\documentclass[conference,11pt]{IEEEtran}

\usepackage{blindtext, graphicx,nomencl,multicol,float,subfigure}

\usepackage{supertabular,booktabs}
\usepackage{import}
\graphicspath{{pictures/}}
\usepackage{graphicx}
\usepackage{fancyhdr}
\usepackage{pst-node,pst-text,pst-3d,pst-char}
\usepackage{array}
\usepackage{siunitx}
\usepackage{todonotes}
\usepackage{epstopdf}

\begin{document}

\title{{\large \vspace*{-3mm} \hspace*{-3mm} 2015 Tenth International Conference on Ecological
Vehicles and Renewable Energies (EVER)}\vspace{5mm}\\
Air-Gap Convection in a \\ Switched Reluctance Machine \vspace*{-7mm}}

\author{\IEEEauthorblockN{Pietro Romanazzi and David A. Howey}
\IEEEauthorblockA{University of Oxford \\ Department of Engineering Science, Parks Road\\ Oxford, OX1 3PJ, UK\\ Email: pietro.romanazzi@eng.ox.ac.uk}}

\maketitle

\begin{abstract}
Switched reluctance machines (SRMs) have recently become popular in the automotive market as they are a good alternative to the permanent magnet machines commonly employed for an electric powertrain. Lumped parameter thermal networks are usually used for thermal analysis of motors due to their low computational cost and relatively accurate results. A critical aspect to be modelled is the rotor-stator air-gap heat transfer, and this is particularly challenging in an SRM due to the salient pole geometry. This work presents firstly a review of the literature including the most relevant correlations for this geometry, and secondly, numerical CFD simulations of air-gap heat transfer for a typical configuration. A new correlation has been derived: $\mathbf{Nu=0.181\ Ta_m^{0.207}}$.
\end{abstract}


\IEEEpeerreviewmaketitle

This paper is part of the ADvanced Electric Powertrain Technology (ADEPT) project which is an EU funded Marie Curie ITN project, grant number 607361. Within ADEPT a virtual and hardware tool are created to assist the design and analysis of future electric propulsion, especially within the context of the paradigm shift from fuel powered combustion engines to alternative energy sources (e.g. fuel cells, solar cells, and batteries) in vehicles like motorbikes, cars, trucks, boats, planes. The design of these high performance, low cost and clean propulsion systems has stipulated an international cooperation of multiple disciplines such as physics, mathematics, electrical engineering, mechanical engineering and specialisms like control engineering and safety. By cooperation of these disciplines in a structured way, the ADEPT program provides a virtual research lab community from labs of European universities and industries \cite{ADEPT}.
\section{Introduction}

\begin{figure}[hb]
\centering
\includegraphics[scale=0.7]{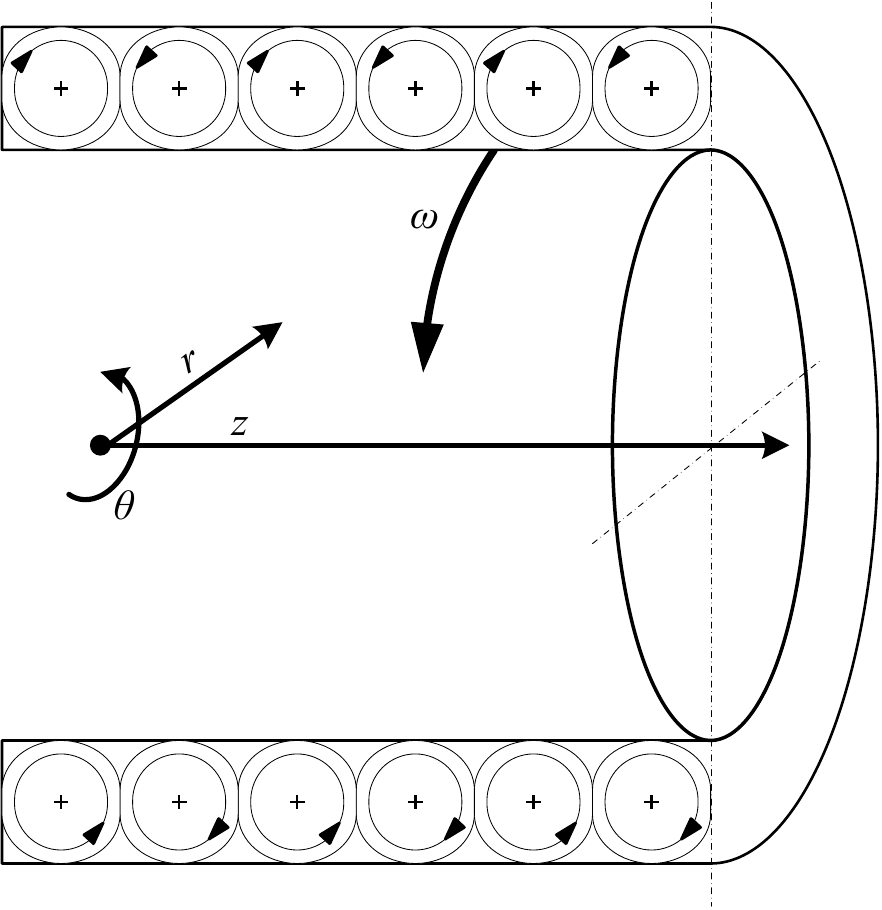} 
\caption{Taylor instability}
\label{fig:taylor}
\end{figure}

Switched reluctance motors (SRMs) are becoming more popular for traction drives \cite{Odin} since they are robust and do not require permanent magnets \cite{Rahman2000}. They have a simple rotor construction and regulated magnetic flux, which can provide a significantly extended speed range. An optimized SRM design can have power densities competitive with a PM machine \cite{Boynov}. However, high torque ripple and associated vibration and noise are possible disadvantages \cite{Santos2014}. In order to achieve high torque densities, water cooling is usually employed for traction motors in electric powertrains. In order to evaluate the cooling effectiveness, thermal analysis must be carried out, modelling all the paths that lead the heat from the active parts to the water jacket and the environment. Lumped parameter thermal networks (LPTN) are one of the most widely used approaches for thermal analysis due to their low computational cost and relatively accurate results. One of the most critical aspects to be modelled is the air-gap heat transfer \cite{Staton2005, Howey2012}, and this is particularly challenging in an SRM due to its salient pole geometry. This heat transfer path is governed by convection and usually the theory developed by Taylor \cite{Taylor1923} is applied. Taylor studied the instability developing in an annulus between two smooth cylinders where the inner one is rotating, finding that pairs of counter-rotating vortices appear in the \(z-r\)  plane (Fig.\ \ref{fig:taylor}) above a particular rotational speed.

\begin{figure}
\centering
\includegraphics[scale=0.5]{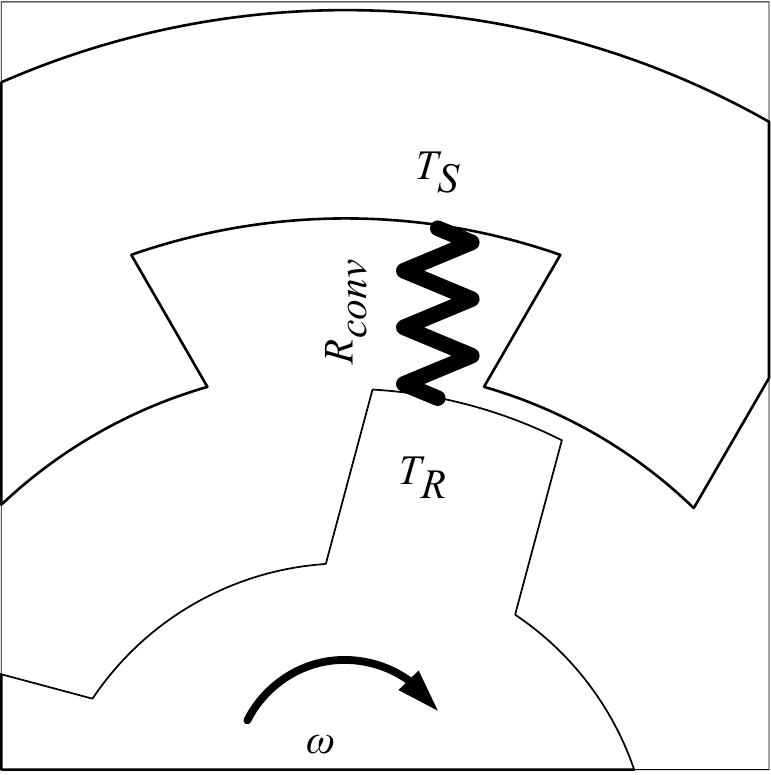} 
\caption{Air-gap equivalent thermal resistance (averaged behaviour)}
\label{fig:resistor}
\end{figure}

As derived in \cite{Howey2012}, assuming the case of adiabatic flow between infinitely long smooth cylinders and very narrow gaps, i.e. \(g/R_i\approx 0\) where $g$ is the gap width and $R_i$ the inner radius of the annulus, Taylor found that the onset of the instability occurs at \(Ta_c=1,697\), where \( Ta \) is a dimensionless group called the Taylor number whose definition will be discussed in the next section. In the case of non-adiabatic flow, with heat transfer between the fluid flow an the boundary walls, Becker and Kaye \cite{Becker1962} derived and proved that the Taylor vortices should appear at \(Ta_c=1,740\).
After Taylor, many other authors explored this particular flow looking for example at the heat transfer between cylinders \cite{Bjorklund1959}, the modes of the instability \cite{Coles1965} or the effect of an additional axial flow \cite{Tachibana1964,Poncet2010,Lancial2014}. A wider review of the problem may be found in \cite{Andereck1992}. In general the following four modes, in order of increasing rotational speed, are observed \cite{Howey2012}:
\begin{itemize}
\item Laminar flow
\item Laminar flow plus Taylor vortices 
\item Turbulent flow plus Taylor vortices
\item Turbulent flow 
\end{itemize}
The presence of salient poles on the rotor side considerably changes the boundary conditions of the problem since the air flow is no longer driven by a continuous surface, and recirculations develop in the slots. This affects the critical value for the onset of instability and the rate of convective heat transfer between the rotor and stator surface, which may increase by up to 50\% \cite{Gazley1958}. Thus, the classical correlations from the literature (e.g. \cite{Becker1962a,Aoki1967,Tachibana1960}) seem not to be ideal for this geometry. We focus on the impact of slots on the heat transfer and flow patterns of an SRM, by first reviewing the most relevant previous works and then undertaking numerical simulations of the airflow using the commercial CFD code STAR-CCM+ v9.06 \cite{STARCCM}. Eventually this will enable the construction of a thermal resistance \(R_{conv}=\frac{1}{A_{h} h}\) representing the air-gap average heat transfer within a LPTN (Fig.\ \ref{fig:resistor}), where $A_h$ is the heat exchanging area and $h$ is the heat transfer coefficient. Such a resistance would be a function of the machine geometry and speed.

\section{Literature Review}
Before starting the review we introduce the dynamic parameters that influence the flow field in the air gap. Since for vehicle traction usually an enclosed motor configuration is applied, no axial air flow is superimposed to the Taylor-Couette flow in the annulus. Thus, the tangential Reynolds number \(Re_{\theta}=\frac{\omega R_i g}{\nu}\) is one of the main parameters that characterize the flow, where $\omega$ is the rotational speed of the internal cylinder and $\nu$ is the kinematic viscosity of air. However, the Taylor number is a more appropriate dimensionless group to use an independent variable in annulus geometries, since it expresses the ratio of the centrifugal to the viscous forces. In our work we will use the so-called modified Taylor number \(Ta_m\) according to the definition introduced by Bouafia \cite{Bouafia1998} to take into account the presence of grooves:
\begin{equation}
Ta_m=\frac{\omega^2 R_m g^3}{\nu^2}~ \frac{1}{F_g}\
\end{equation}
Where \(R_m\) is the mean radius defined as \(R_m=b/ln((R_i+b)/R_i)\) with \(b=A/2\pi R_i\), and \(Fg\) is a geometric factor:
\begin{equation}
Fg=\frac{\pi^4}{1967 P}~\frac{R_i+R_o}{2 R_i} 
\end{equation}
and 
\begin{equation}
P=0.0571(1-0652\frac{g}{R_i})+0.00056(1-0.652\frac{g}{R_i})^{-1}
\end{equation}
The Nusselt number is the dependent variable; this characterizes the heat transfer by convection and the following definition has been used:
\begin{equation}
Nu=\frac{q~g}{k~(T_R-T_S)}=\frac{h~g}{k}
\end{equation}
where $q$ is the specific heat flux, $T_R$ and $T_S$ are the surface temperatures of the rotor and stator side respectively and $k$ is the thermal conductivity of air. We refer to rotor and stator sides using the subscripts R and S respectively. Nusselt number will be calculated as the average value over the entire internal surface of the annulus. The thermal conductivity \(k\) is taken as the mean of the values at the two surface temperatures.

Throughout the literature there are a number of different definitions for these parameters \cite{Fenot2011}. Any system of definitions may be used as long as consistency is maintained, however for the case of slotted surfaces we use the system of \cite{Bouafia1998}, as previously defined. Based on the various experimental geometries presented in Tab. \ref{tab:geometry}, we have re-expressed the correlations using a consistent set of definitions of $Ta$ and $Nu$ in order to compare like with like. The values are also compared with the case of smooth cylinders using the correlation by Becker and Kaye \cite{Becker1962}:
\begin{itemize}
  \item[] $Nu=1$ for $Ta_m<Ta_c$
  \item[] $Nu=0.064\ {Ta_m}^{0.367}$ for $Ta_c<Ta_m<10^4$
  \item[] $Nu=0.205\ {Ta_m}^{0.241}$ for $10^4<Ta_m<10^6$
\end{itemize}

\setlength{\extrarowheight}{2pt}
\begin{table}[ht]
\centering{\caption{Geometry Dimensions From Literature Related To Fig.\ \ref{fig:geometry} - [$\mathrm{mm}$]}
\label{tab:geometry}
\begin{tabular}{l c|c|c|c|c|c}

&\multicolumn{1}{c}{\textbf{Bouafia}}&\multicolumn{2}{c}{\textbf{Hayase}}&\multicolumn{2}{c}{\textbf{Gazley}}&\multicolumn{1}{c}{\textbf{Hwang}} \\ 

&\multicolumn{1}{c}{\cite{Bouafia1998}\cite{Bouafia1999}}&\multicolumn{2}{c}{\cite{Hayase90}\cite{Hayase92}}&\multicolumn{2}{c}{\cite{Gazley1958}}&\multicolumn{1}{c}{\cite{Hwang1990}}\\ [4pt]

\toprule Case & n.a.& B& C& S1-R2& S2-R2& n.a.  \\ [3pt]
\hline $g$& $5$ & $0.05$& $0.05$& $0.5$& $0.5$& $2$\\ [3pt]
\hline $R_i$& $140$ & $0.95$& $0.95$& $63.3$& $63.3$& $34$\\ [3pt] 
\hline $L_{RW}$& $-$ & $-$&$-$& $0.1$& $0.1$& $\delta\frac{2\pi R_i}{16}$\\ [3pt]
\hline $L_{RD}$& $-$ & $0.2$& $-$& $0.76$& $0.76$& $2$\\ [3pt]
\hline $\alpha_R$& $-$ & $10^{\circ}$& $-$& $0^\circ$& $0^\circ$& not given\\ [3pt]
\hline $N_R$& $0$ & $12$& $0$& $36$& $36$& $16$\\ [3pt]
\hline $R_o$& $145$ & $1.0$& $1.0$& $63.8$& $63.8$& $36$\\ [3pt] 
\hline $L_{SW}$& $8.3$ & $-$&$-$& $-$& $4.6$& $-$\\ [3pt]
\hline $L_{SD}$& $15$ & $-$& $0.2$& $-$& $1.5$& $-$\\ [3pt]
\hline $\alpha_S$& $0^\circ$ & $-$& $10^{\circ}$& $-$& $0^\circ$& $-$\\ [3pt]
\hline $N_S$& $48$ & $0$& $12$& $0$& $23$& $0$\\ [3pt]

\end{tabular}}
\vspace{-3mm}
\end{table}

\begin{figure}
\centering
\includegraphics[scale=2,keepaspectratio]{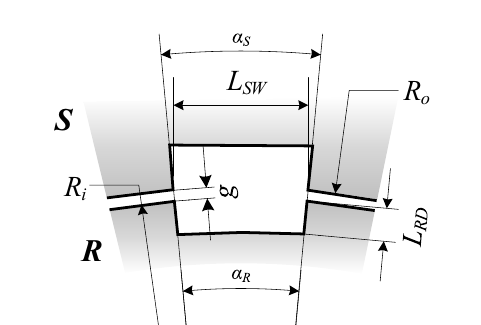}
\caption{Reference geometry}
\label{fig:geometry}
\end{figure}

It is worth noting that only Hayase \cite{Hayase90,Hayase92} numerically simulated a domain geometrically similar to an SRM rotor. However, the simulations were performed considering a geometry with an higher ratio of the air gap to the slot height compared to a typical SRM configuration. Also Srinivas \cite{Srinivas2004, Srinivas2005} ran some CFD simulations of the SRM airgap, although assuming a 2-D and incompressible domain and without reporting any detail about the heat transfer.

\begin{figure*}%[!ht]
  \includegraphics[width=7in,keepaspectratio]{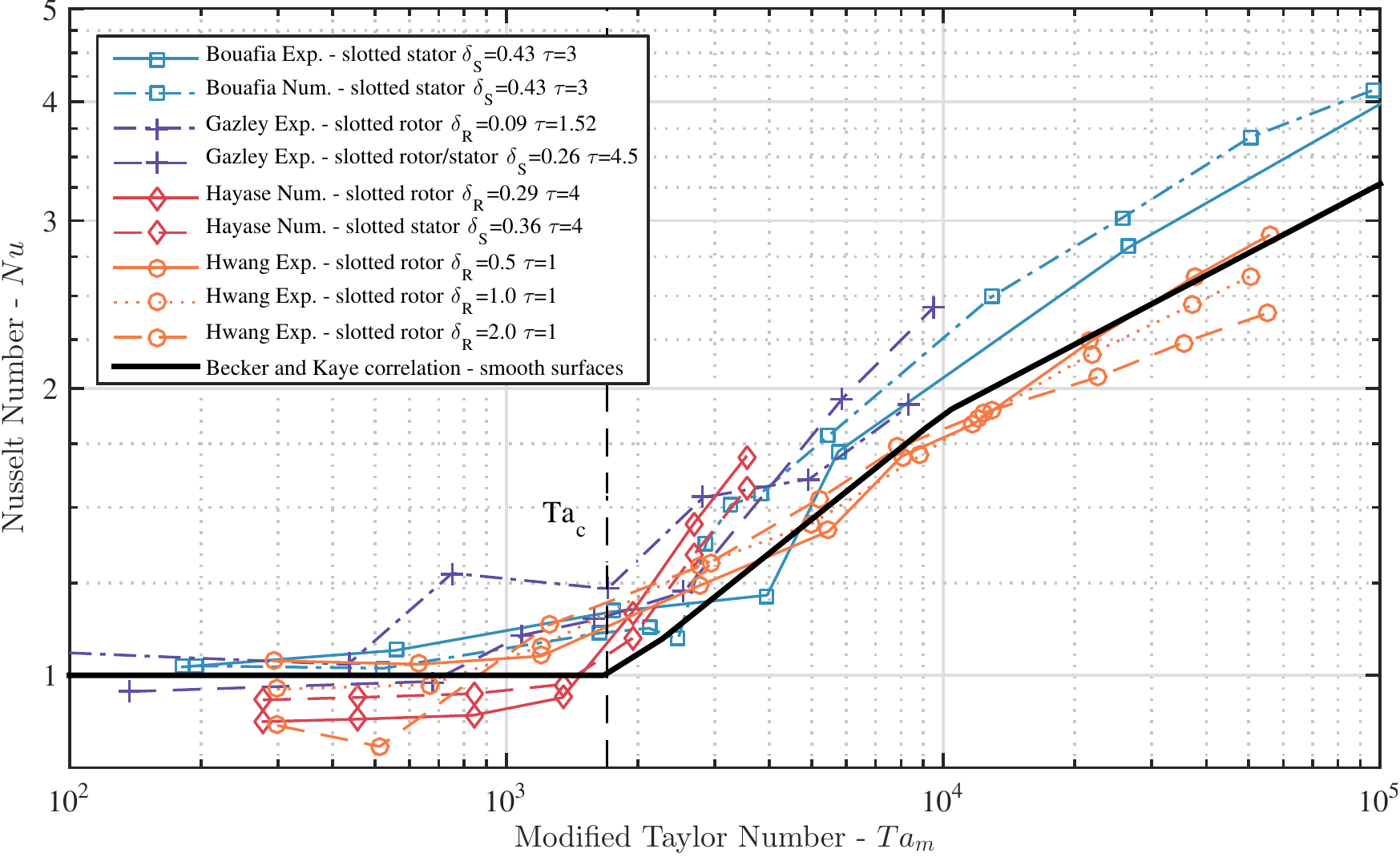}
  \caption{Heat transfer vs. Taylor number for a range of previous experimental and numerical studies}
  \label{fig:matlabplot}
\end{figure*}

Fig.\ \ref{fig:matlabplot} shows that in the case of slotted surfaces in the laminar regime \(Nu\) is in general less than 1 and increasing at a small constant rate. However, keeping in consideration the ratios \(\tau=\frac{L_D}{g}\) and \(\delta=\frac{L_W}{2 \pi R}\frac{1}{N}\), where \(N\) is the number of slots, it seems that the larger the slot, the lower \(Nu\). This has also been indicated by Tachibana and Fukui \cite{Tachibana1960} who in their analysis also tested the influence of gap sizes at low rotational speeds. 

As the rotational speed increases beyond a critical point the heat transfer rate suddenly rises since vortices appear which promote fluid recirculation. However, the results from the literature give some uncertainty for what the critical \(Ta_m\) is. For Hwang \cite{Hwang1990}, Hayase \cite{Hayase92} and Gazley \cite{Gazley1958} the instability onset seems to take place for \(Ta_m < Ta_c\). Whereas according to the data from Bouafia \cite{Bouafia1998,Bouafia1999}, the vortices appear for higher values of modified Taylor number, and in particular the critical \(Ta_m=3,900\) is measured.  
This discrepancy is mainly due to the various configurations studied which differ in geometry and whether the rotor is the inner or outer side of the annulus.

An interesting work which has not been shown in Fig.\ \ref{fig:matlabplot} is Gardiner and Sabersky \cite{Gardiner1978}, where the values lie out of the main area of data due to the fact that the fluid employed was water (Prandtl number \(Pr=4.5\)). However, their measurements show again a similar behaviour of the heat transfer rate with a sudden increase at \(Ta_m \approx 3\times 10^{3}\). 
More recently the heat transfer in an annulus between coaxial cylinders with slots had been studied \cite{Poncet2010,Fenot2013,Lancial2014}. However, these works analyse a Taylor-Couette-Poiseuille flow, meaning that an axial air flux is superimposed across the Taylor-Couette flow generated by the rotation of the internal or external cylinder.

\section{Software validation}
\begin{figure}[ht]
  \includegraphics[width=3.5in,keepaspectratio]{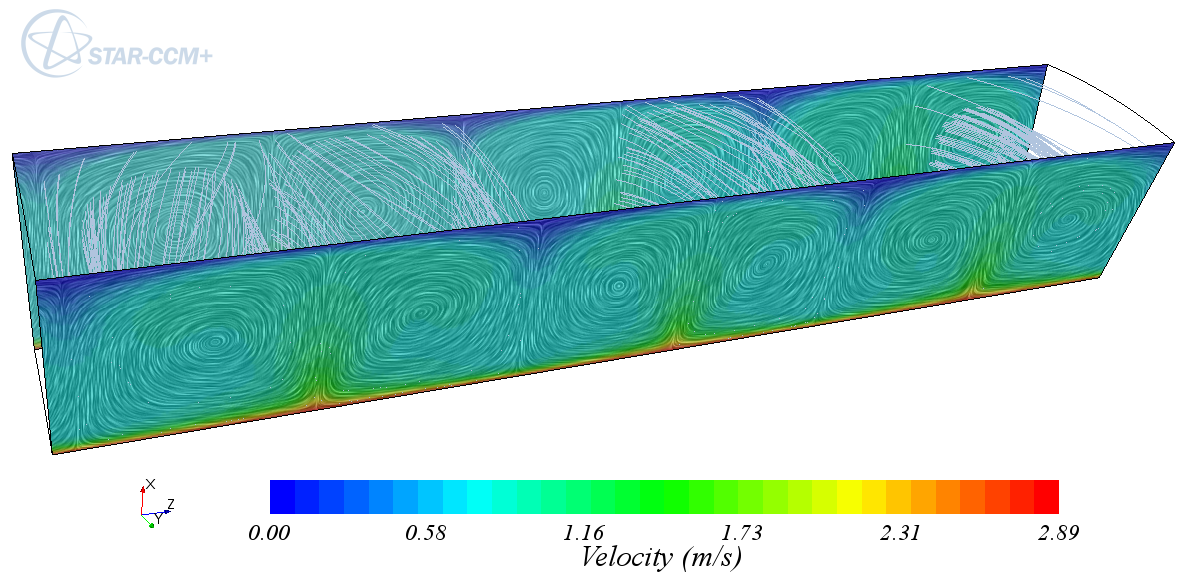}
  \caption{Taylor-Couette flow pattern \(\eta=0.75\)}
  \label{fig:vorticini}
\end{figure}

Computational fluid dynamics simulations were undertaken using CD-adapco's STAR-CCM+. First, the baseline scenario of heat transfer between two smooth cylinders with the inner one rotating and heated was simulated. By assuming circumferential periodicity, the calculations were confined to 30$^{\circ}$ in the $r - \theta$ plane as assumed by Hayase \cite{Hayase90,Hayase92}. The axial length of the domain was set to allow an even number of vortices to develop \cite{Donnelly1965}. Additionally, to prevent end effects, a periodic boundary condition was applied to the axial ends to create an infinitely long annulus. The no-slip condition was applied to the outer and inner walls with the inner wall rotating at a constant speed. Constant temperatures of 60\ \(^\circ\)C and 35\ \(^\circ\)C were imposed to the inner and outer walls respectively. In Fig.\ \ref{fig:vorticini} the model is presented for the case of \(\eta=\frac{R_i}{R_o}=0.75\) and rotational speed large enough to let the instability grow, showing the velocity field characterised by the expected counter-rotating vortices.  
This steady state solution was achieved using the \(k-\epsilon\) turbulence model and the two-layer all \(y^+\) wall treatment \cite{STARCCM}, starting the simulation from the fluid at rest. As far as the fluid domain, air was assumed to be an ideal gas with constant specific heat \(C_{air}\), thermal conductivity \(k_{air}\) and dynamic viscosity \(\mu=\nu\rho_{air}\), where \(\rho_{air}\) is the air density. All values were calculated at the average temperature between the rotor and the stator surfaces. When it comes to simulating very narrow gaps, i.e.\ \(\eta \approx 0.99\) as in the case of the SRM and electrical machines in general, this turbulence model was unable to reproduce the instability correctly. As suggested in \cite{Chandra}, the large eddy simulation (LES) approach has therefore been applied due to its ability to resolve smaller structures compared to the models based on Reynolds averaged Navier-Stokes equations. The WALE (wall-adapting local eddy-viscosity) subgrid scale model was chosen to estimate the subgrid scale viscosity along with the two-layer all \(y^+\) wall treatment to obtain accurate results with a reduced computational time \cite{STARCCM}. The simulations were then run with this unsteady solver applying the same surface temperatures, no-slip conditions and periodic boundaries. Convergence was assumed to be reached when the value of \(Nu\) calculated as the average over the rotating surface settled to a constant value, which happened after 0.1\ $\mathrm{s}$ simulation time.

\begin{figure}
  \includegraphics[keepaspectratio]{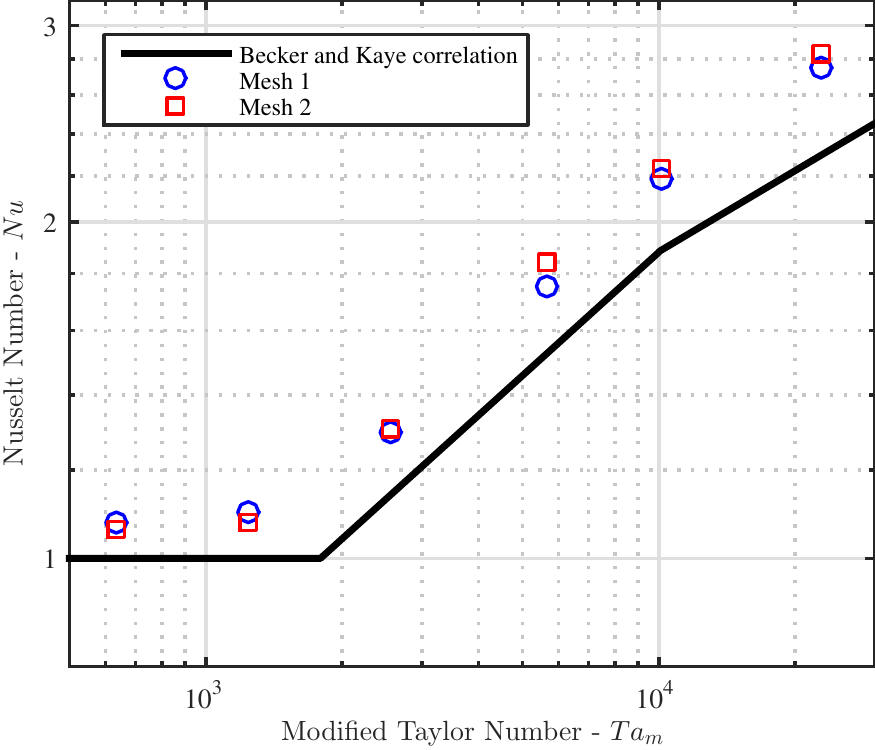}
  \caption{LES model validation and mesh comparison}
  \label{fig:validation}
\end{figure}

In Fig.\ \ref{fig:validation} we present a comparison between the numerical simulations using the LES model at two different mesh sizes, and the correlation from Becker and Kaye \cite{Becker1962}. Both the meshes were built with the STAR-CCM+ using polyhedral elements: Mesh 1 consist of 262,406 elements and Mesh 2 of 512,365; the higher number of elements effected only the bulk area between the prism layers. To correctly resolve the fields near to the boundaries a finer mesh consisting of 20 prism layers with a growing factor of 1.2 was applied, with the first layer height of 1.2$\times10^{-6}$ m. This mesh resolution allowed always a \(y^+\) value below 1. The results indicate that the LES model is able to reproduce the increase of heat transfer due to the onset of the instability. The results are fairly mesh independent as shown in Fig.\  \ref{fig:validation} which gives average $Nu$ of the numerical solutions over about two rotations starting after the fluid flow settled to an almost steady pattern. 

\section{Results and analysis}
\subsection{Modelling the SRM}
\begin{figure}
\includegraphics[scale=0.5]{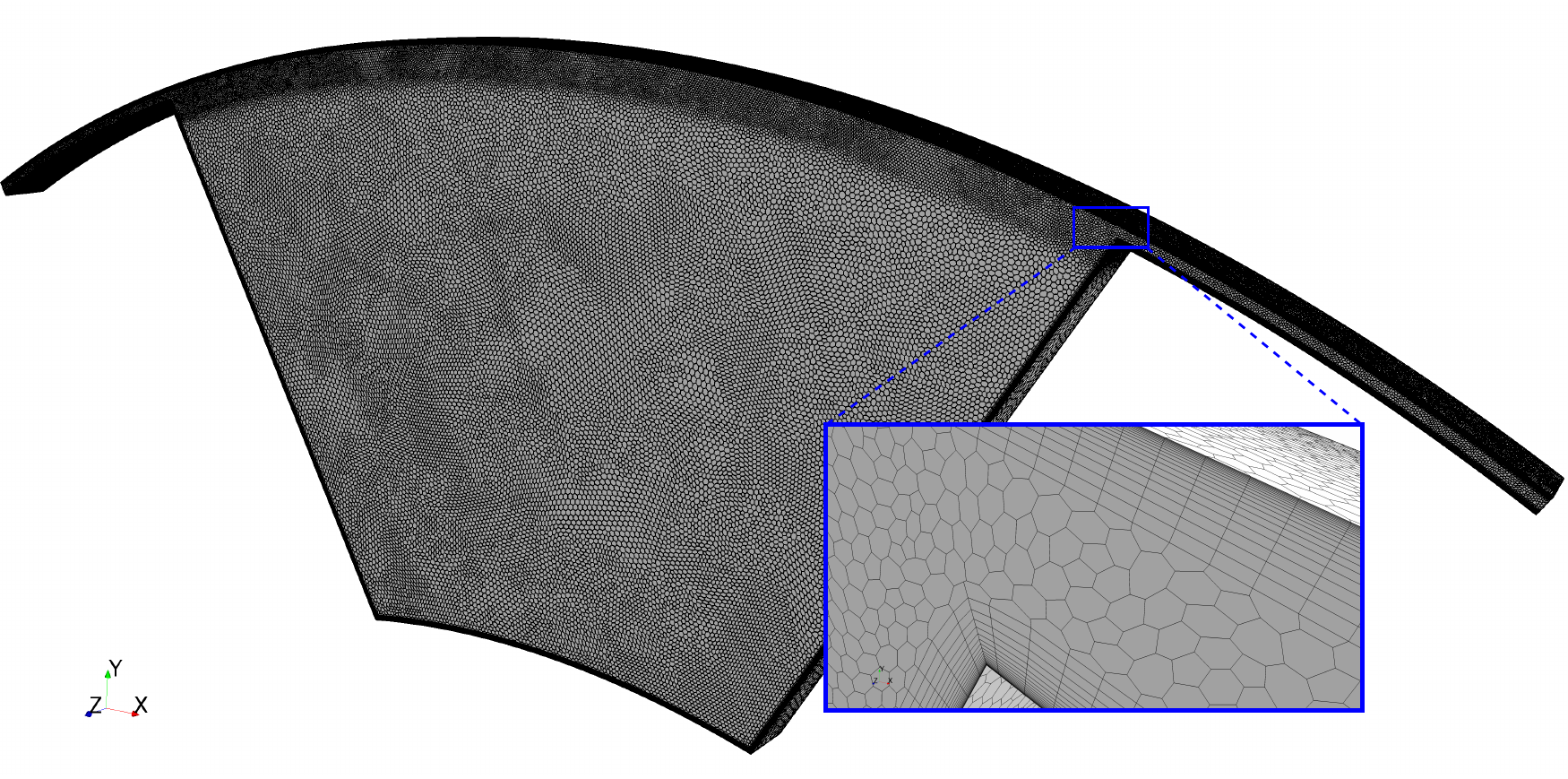} 
\caption{SRM 8/6 meshed model}
\label{fig:mesh}
\end{figure}

The case of an 8/6 SRM (8 salient poles on the stator side and 6 on the rotor side) was  considered. The geometry details were taken from \cite{Rahman2000} since we are mainly interested in SRMs for electric vehicle applications. These dimensions, presented in Table\ \ref{tab:SRMgeometry},  are in line with the general design rules given in \cite{Miller}. For this research we modelled the stator surface to be smooth (assuming the presence of windings and slot liner).
\begin{table}
\centering
\caption{Srm Model Dimensions}
\label{tab:SRMgeometry}
\begin{tabular}{l r l}
\toprule
stator inner radius & $95.95$ & $mm$\\
rotor outer radius & $95.00$ & $mm$\\
air gap width & $0.95$ & $mm$\\
rotor pole height & $33.5$ & $mm$\\ 
rotor pole arc & $23^\circ$ & -\\
number of slots & $6$ & -\\
\bottomrule
\end{tabular}
\end{table}
Due to the periodicity of the structure, only $60^\circ$ of the machine has been studied, applying a periodic boundary condition on the tangential surfaces. Again, a periodic boundary condition has been applied at the axial ends to mimic an infinitely long annulus and prevent end effects. The length of the model was set to accommodate a couple of counter-rotating vortices. The no-slip condition was set on the stator and rotor surfaces which are assumed to be heated at a constant temperature of 60\ \(^\circ\)C and 35\ \(^\circ\)C respectively. To simulate the motion, the reference frame was set to rotate at a specific $\omega^*$, the outer cylinder was spinning at $\omega_S=-\omega^*$ and the inner wall was at $\omega_R=0$ relative to the rotating frame.
The mesh, consisting of 1,450,394 cells, is characterized by two different scales: one finer, keeping the Mesh 1 set up which correctly simulated Taylor vortices, and one coarser for the slot area where larger eddies appear. As shown in Fig.\ \ref{fig:mesh}, 20 prism layers was built next to the wall to ensure \(y^+\) to be lower than 1.

\subsection{Results}

\begin{figure}[b]
\centering
  \includegraphics[width=3.5in,keepaspectratio]{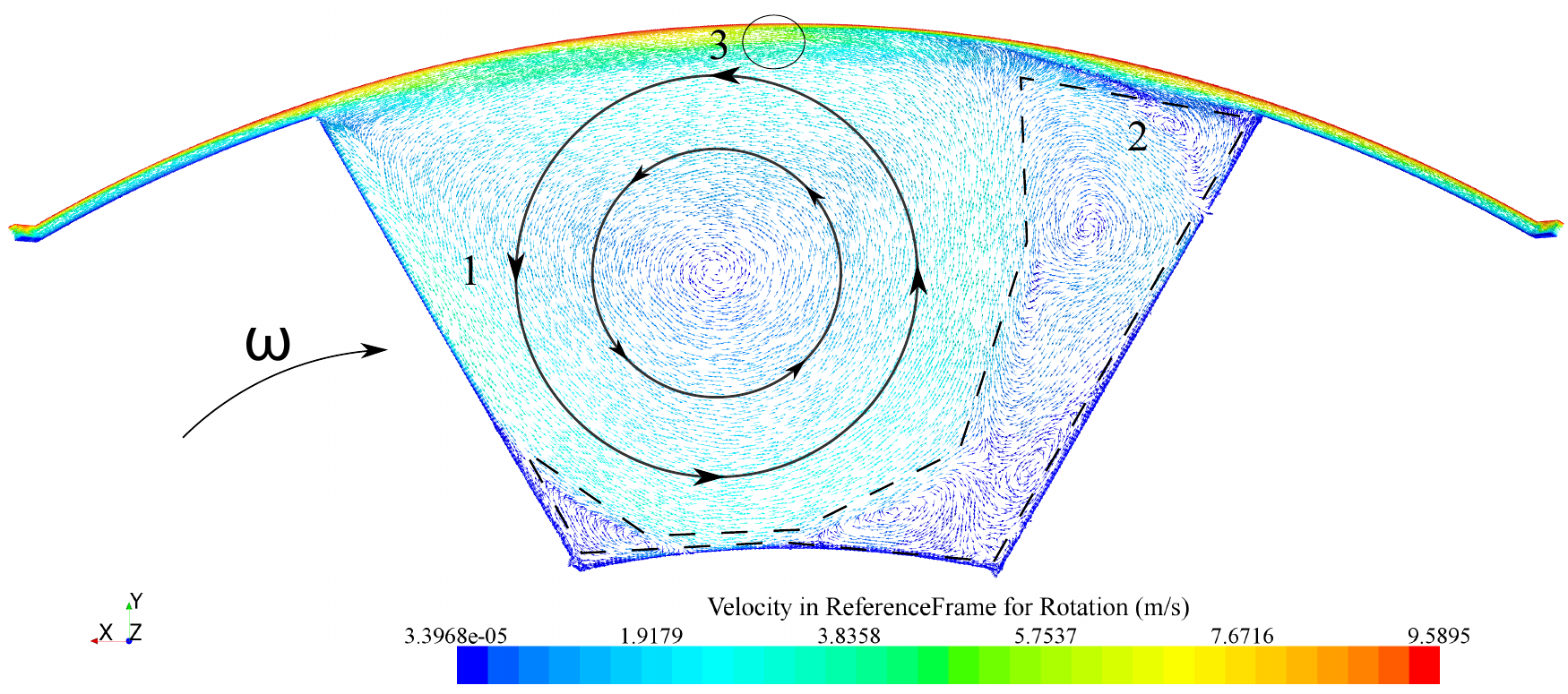}
  \caption{Velocity field with the reference to the rotor at $Ta_m=2,802$}
  \label{fig:vel}
\end{figure}
\begin{figure}
  \includegraphics[width=3.5in,keepaspectratio]{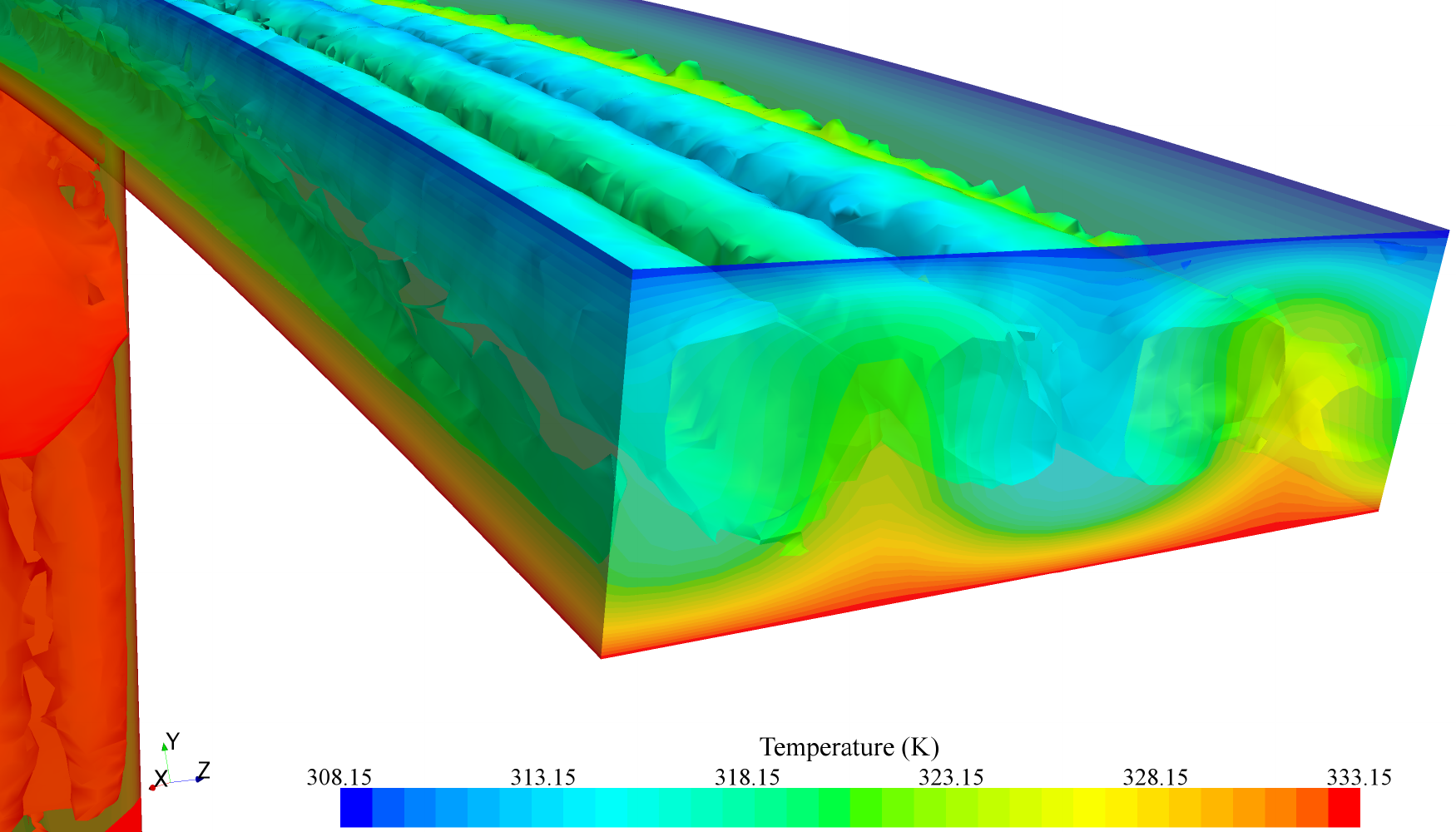}
  \caption{Visualisation of Taylor vortices in the SRM airgap, $Ta_m=11,208$}
  \label{fig:taylorvor}
\end{figure}

A number of rotational speeds were tested for the SRM domain, running each simulation using the previous solution as initial condition. In Fig.\ \ref{fig:vel} the velocity field relative to the rotating frame is presented in the \(r-\theta\) at \(Ta_m=2,802\). The presence of the slot leads the flow to separate, generating some vortices within the slot.  
The simulations showed that in the slot a dominant vortex, marked with the number 1 in Fig.\ \ref{fig:vel}, grows with the increasing rotor speed $\omega$. In the ``2" area various minor vortices develop, which tend to split and flatten on the trailing and bottom edges of the slot with $\omega$. Although in the slot the flow has a prevailing 2D behaviour, in the area marked with number 3 a couple of counter-rotating vortices in the \(z-r\) plane were noticed, growing from low values of \(Ta_m\) ($\approx 300$).
The presence of recirculation flows in the slot seem to delay the appearance of Taylor vortices. The onset of the instability was not recorded at $Ta_c$ but at higher values of $Ta_m$. Moreover, the counter rotating vortices were not continuous all around the circumference, but the simulations showed them to grow only in the gap between the rotor pole tip and the stator. This in turn affected the stability of the structures in the slot. Fig.\ \ref{fig:taylorvor} shows the effect of the Taylor vortices on the temperature field. 

\begin{figure}
  \includegraphics[width=3.5in,keepaspectratio]{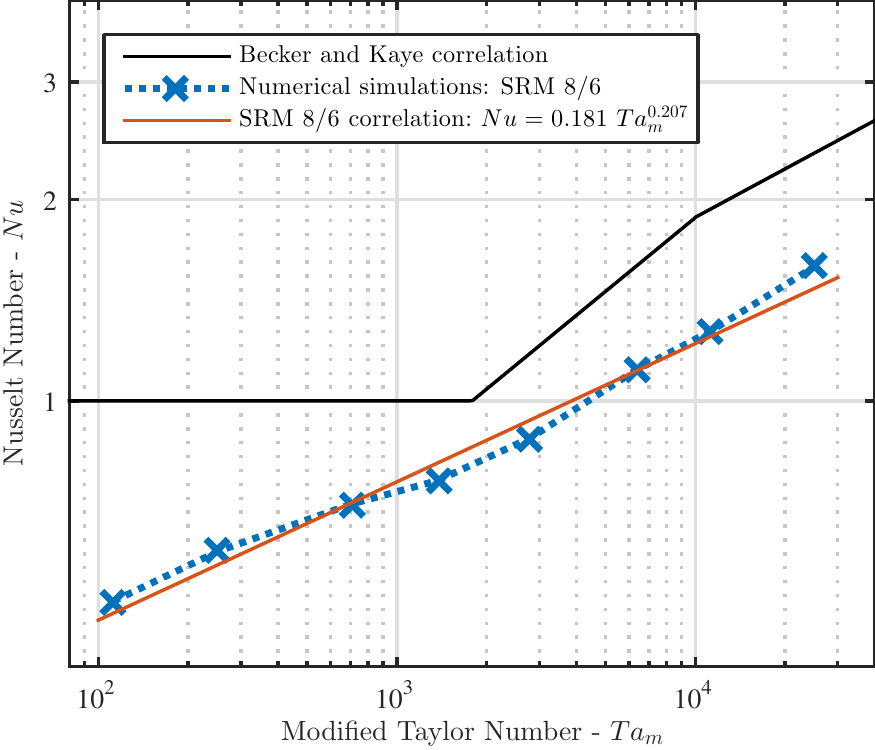}
  \caption{Simulated heat transfer vs.\ Taylor number for an SRM 8/6}
  \label{fig:final}
\end{figure}

The calculated heat transfer rate is presented in Fig.\ \ref{fig:final}. Due to the unsteady solver, the value of \(Nu\) is calculated averaging the solution over about three revolutions after the transition was resolved ($\approx 1$ rotation). We can see that the heat transfer for low value of $Ta_m$ is below unity $Nu$ and increases with $\omega$ without experiencing the expected knee indicating a regime change. The surfaces bordering the slot are exposed to recirculation which reduces the thermal gradient and accordingly the heat transfer rate. As the ratio of such surfaces to the surface on the tip of the pole is higher than 1, the effect of the Taylor instability on the overall heat transfer rate is not noticed. 

\section{Conclusions and future work}
In this work the case of heat transfer in the air gap of an SRM was tested. First, our CFD simulations run with the unsteady LES solver embedded in the commercial software STAR-CCM+, showed good agreement between the case of heat transfer in smooth narrow gaps. For the SRM configuration, the CFD results highlighted that with the presence of SRM rotor slots the instability appears for higher values of \(Ta_m\), in agreement with what was found in previous works. It is important to notice that through the whole tested range of rotational speeds, \(Nu\) never exceeded the correlation given by Becker and Kaye \cite{Becker1962}. A correlation has been derived by least-square fitting the data: $Nu=0.181\ Ta_m^{0.207}$ for $10^2<Ta_m<3\times10^4$.
Future work will aim at analysing the heat transfer at very high rotational speeds, i.e. $10,000$ rpm  which correspond to \(Ta_m \propto 10^5\), and exploring the effect of different slot dimensions (e.g. SRM 6/4).

\section*{Acknowledgments}
The authors would like to thank the European Union for their funding to this research (FP7 ITN Project 607361 ADEPT) and CD-adapco and Dr Brian Tang for their support.

\bibliographystyle{IEEEtran}
\bibliography{library}

\end{document}